\documentclass[a4paper,11pt]{article}
\usepackage{amsmath,amsthm,amssymb,mathrsfs,graphicx,color}
\usepackage[titletoc,title]{appendix}
\usepackage{comment} 
\newcommand{\nc}{\newcommand}
\nc{\rnc}{\renewcommand}
\nc{\nn}{\nonumber}
\nc{\der}{{\partial}}
\rnc{\Im}{{\textrm{Im}\,}}
\rnc{\Re}{{\textrm{Re}\,}}

\nc{\rmi}{{\rm i}}
\nc{\rme}{{\rm e}}
\nc{\rmA}{{\rm A}}
\nc{\rmB}{{\rm B}}
\nc{\rmC}{{\rm C}}
\nc{\rmD}{{\rm D}}
\nc{\rma}{{\rm a}}
\nc{\rmb}{{\rm b}}
\nc{\rmc}{{\rm d}}
\nc{\rmd}{{\rm d}}
\nc{\bra}{\langle}
\nc{\ket}{\rangle}

\nc{\tcr}{\textcolor{red}}
\nc{\tcb}{\textcolor{blue}}

\nc{\End}{\mathrm{End}}

\theoremstyle{definition}

\numberwithin{equation}{section}

\begin{document}

\title{Quantum random access memory via quantum walk}

\author{
Ryo Asaka\thanks{E-mail: 1219502@ed.tus.ac.jp}, \,
Kazumitsu Sakai\thanks{E-mail: k.sakai@rs.tus.ac.jp} \, and
Ryoko Yahagi \thanks{E-mail: yahagi@rs.tus.ac.jp}
\\\\
\textit{Department of Physics,
Tokyo University of Science,}\\
 \textit{Kagurazaka 1-3, Shinjuku-ku, Tokyo, 162-8601, Japan} \\
\\\\
\\
}

\date{August 31, 2020}

\maketitle

\begin{abstract}
A novel concept of quantum random access memory (qRAM) employing a quantum 
walk is provided. Our qRAM relies on a bucket brigade scheme to access the 
memory cells. Introducing a bucket with chirality {\it left} and {\it right} 
as a quantum walker, and considering its quantum motion on a full binary tree,
we can efficiently deliver the bucket to the designated memory cells, and 
fill the bucket with the desired information in the form of quantum superposition 
states. Our procedure has several advantages. First, we do not need to place
any quantum devices at the nodes of the binary tree, and hence in our qRAM 
architecture, the cost to maintain the coherence can be significantly reduced.
Second, our scheme is fully parallelized.  Consequently, only  
$O(n)$ steps are required to access and 
retrieve $O(2^n)$ data in the form of quantum superposition states. 
Finally, the simplicity of our procedure may allow the design of 
qRAM with simpler structures.
\end{abstract}

%
%%%%%%%%%%%%%%%%%%%%%%%%%%%%%%%%%%%%%%%%%%%%%%%%%%%%%%%%%%%%
\section{Introduction}
%%%%%%%%%%%%%%%%%%%%%%%%%%%%%%%%%%%%%%%%%%%%%%%%%%%%%%%%%%%%
%
The development of an efficient procedure to retrieve classical/quantum data from a  
database and transform them into a quantum superposition state is one of the most 
fundamental issues for a practical realization of quantum information processing.  
Quantum random access memory (qRAM) that stores information and permits queries in 
superposition may play a pivotal role in a substantial speedup of quantum algorithms 
for data analysis  \cite{Grover,HHL,LMR}, including applications to machine 
learning for big data \cite{LMR2,RML,Nature,SFP,BDLK}.

qRAM is a quantum analog of classical RAM. Provided a superposition of 
addresses $\sum_{a}|a\ket$ ($a\in\mathbb{Z}_{\ge 0}$) as input, qRAM accesses
the $a$th cell in the memory array, where  classical 
information $|x^{(a)}\ket$  is stored, and 
outputs a superposition of $|x^{(a)}\ket$'s correlated with the addresses.
Note here that qRAM is possible to store either classical information
(i.e. each $|x^{(a)}\ket$ does not consist of any superposition of states) 
or quantum information(i.e. each 
$|x^{(a)}\ket$ is an arbitrary superposition of states). In this 
paper, we restrict ourselves to the classical case, and  assume
that the addresses $|a\ket$ and the data $|x^{(a)}\ket$ ($a, x^{(a)}\in\mathbb{Z}_{\ge 0}$)
consist of $n$ and $m$ qubits, respectively.

More precisely, qRAM is defined as
\begin{align}
\text{qRAM}:\,\,&
\sum_{a}|a \ket_A  |0\ket_D \mapsto \sum_a |a \ket_A |x^{(a)}\ket_D, \nn \\
&|a\ket_A\in\mathbb{C}^{2n},\,\, 
|x^{(a)}\ket_D\in \mathbb{C}^{2m} \quad (n,m\in\mathbb{Z}_{\ge 0}).
\label{qram-concept}
\end{align}
where $A$ and $D$, respectively, 
denote quantum analogs of  an address register (input register) and a data
register (output register).
 Namely, qRAM \eqref{qram-concept} is a quantum device 
consisting of (i) a routing scheme to access the designated memory cells, 
(ii) a querying scheme to retrieve the data stored in the cells, and (iii) 
an output scheme to encode the data into a superposition of quantum states.

\begin{figure}
\centering
\includegraphics[width=0.9\textwidth]{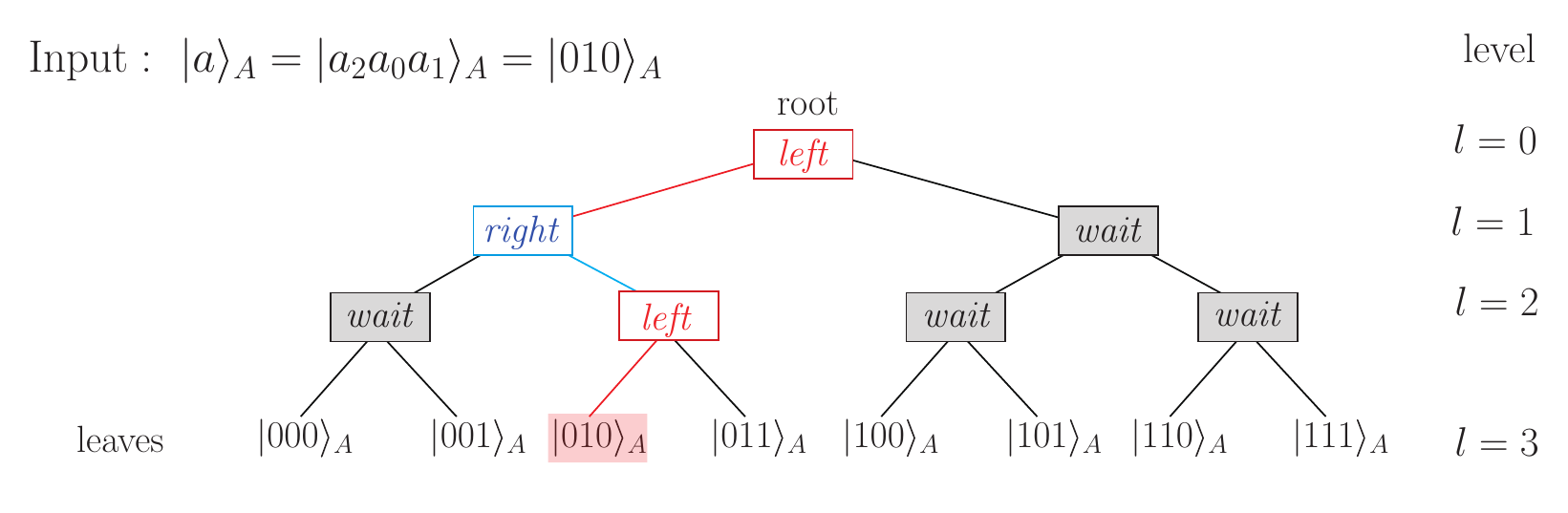}
\caption{A bucket brigade scheme on the full binary tree with depth
$n=3$.  The qutrits are equipped at each node. To route to the memory 
cell $|010\ket_A$, one must activate the three qutrits as in the figure.}
\label{bucket}
\end{figure}
As a quantum routing scheme, a notable idea, the ``bucket brigade" scheme
has been proposed by Giovannetti, Lloyd and  Maccone (GLM)
\cite{GLM-PRL,GLM08} to overcome difficulties associated with
the conventional fanout  scheme commonly implemented in classical RAM  
(see \cite{JB,SS}, for instance). 
The bucket brigade architecture employs a perfect binary tree with $O(N)$ 
($N=2^{n}$) nodes routing a signal from the 
root down to one of the $N$ leaves interpreted as the memory cells (see Fig.~\ref{bucket}
for $n=3$.). 
Let $|a_{n-1}\cdots a_0\ket$ ($a_l\in\{0,1\}$; $0\le l\le n-1$) be the 
binary representation of the address of the cell.
The value of $a_l$ indicates the route from a parent node at the $l$th level to 
one of the two children nodes at the $(l+1)$th level. 
For instance,  the left (resp. right) child is chosen if $a_l=0$ (resp. $a_l=1$). 
In consequence, each of the $N$ possible values of the address register uniquely 
determines a  path in the binary tree.

In the original GLM architecture, a qutrit (i.e. a three-level quantum 
system with energy levels labeled by {\it wait}, {\it left} and {\it right})
is allocated at each node, and all the qutrits are initialized to be {\it wait}
state. 
The value $a_l$ is sequentially delivered from the root to a node at the 
$l$th level, and activates the qutrit to  {\it left} ({\it right})
if $a_l=0$ ($a_l=1$)   to route the following  $a_{l+1}$ 
to  one of the two subsequent nodes. After $O(n^2)$ time steps,
a unique path from the root to the designated memory cell is assigned
as depicted in Fig~\ref{bucket} for $n=3$ and $|a\ket_A=|010\ket_A$.
Remarkably, only  $n$ qutrits are activated, which is exponentially less than 
that for the traditional fanout architecture, where $N$ quantum 
switches are necessary to be activated. Namely, the GLM bucket brigade
architecture has a significant advantage in maintaining quantum coherence. 

A quantum signal (the so-called  quantum bus) 
can follow the path to the desired memory cell through the
activated qutrits, retrieve 
information  stored in the cell, and goes back to the root
along the path. Finally, resetting the activated qutrits to be 
the initialized state {\it wait} one by one starting from the
last level of the tree, one obtains the output as in
the r.h.s. of \eqref{qram-concept}. That is the GLM bucket 
brigade qRAM.
The GLM scheme has been improved, and concretely implemented 
into quantum circuits as in \cite{HXZJW,AGJMS,MGM,POB}. (Note that a 
different concept of qRAM  without relying on any routing 
schemes has recently been developed in \cite{PPR}.)

This paper provides a novel concept of qRAM, 
which employs a discrete-time quantum walk as a bucket brigade scheme. 
A quantum walk is a quantum motion of a particle (interpreted as a bucket) 
possessing  chirality {\it left} and {\it right} \cite{ADZ,VA}. 
The quantum bucket with 
chirality {\it left} (resp. {\it right}) on a parent node moves 
to the left (resp. right)  child node. Each scheme of qRAM can
actually be realized by quantum motions of multiple quantum walkers.
Our procedure has several advantages.
First, because we do not need to equip any quantum devices at the totally $O(n 2^n)$ nodes
on the binary tree,  the cost to maintain the coherence can be reduced. 
Consequently, our qRAM may 
be much more robust against 
the decoherence arising from noises at the nodes.  
Second, each procedure is fully
parallelized. As a result, only  $O(n)$ steps are required to access and
retrieve data in the form of quantum superposition states.
Finally, since the three schemes required in qRAM are entirely independent of each other, 
the architecture of qRAM  can be simplified.

The layout of this paper is as follows. In the subsequent section, we 
introduce a quantum walk on a binary tree. qRAM utilizing
the quantum walk is constructed in Sec.~3. In Sec.~4, we give some specific 
examples of how to implement our qRAM scheme using multiple quantum walkers.
The last section is devoted to a summary.

%
%%%%%%%%%%%%%%%%%%%%%%%%%%%%%%%%%%%%%%%%%%%%%%%%%%%%%%%%%%%%
\section{Quantum walk on a binary tree}
%%%%%%%%%%%%%%%%%%%%%%%%%%%%%%%%%%%%%%%%%%%%%%%%%%%%%%%%%%%%
%
Quantum walks, which are the quantum counterparts of classical random
walks, are defined as a class of unitary time-evolutions on graphs. In contrast to 
the classical random walks, the randomness comes from a superposition of 
quantum states and its time evolution. Here, let us introduce a discrete-time quantum walk on a full binary tree. A quantum particle with chirality $|0\ket_C$ ({\it left}) and
 $|1\ket_C$ ({\it right}) may be interpreted as a quantum ``bucket".  A bucket with chirality
$|0\ket_C$ (resp. $|1\ket_C$) deviates left (resp. right) at each node of the binary tree,
which is in contrast to a bucket in the GLM architecture, where the route 
is determined by activating the qutrit equipped at each node (see the previous section).
In consequence, we do not need to equip any quantum devices at the $O(n 2^n)$ nodes, 
which is one of the main advantages in our scheme.

\begin{figure}
\centering
\includegraphics[width=0.8\textwidth]{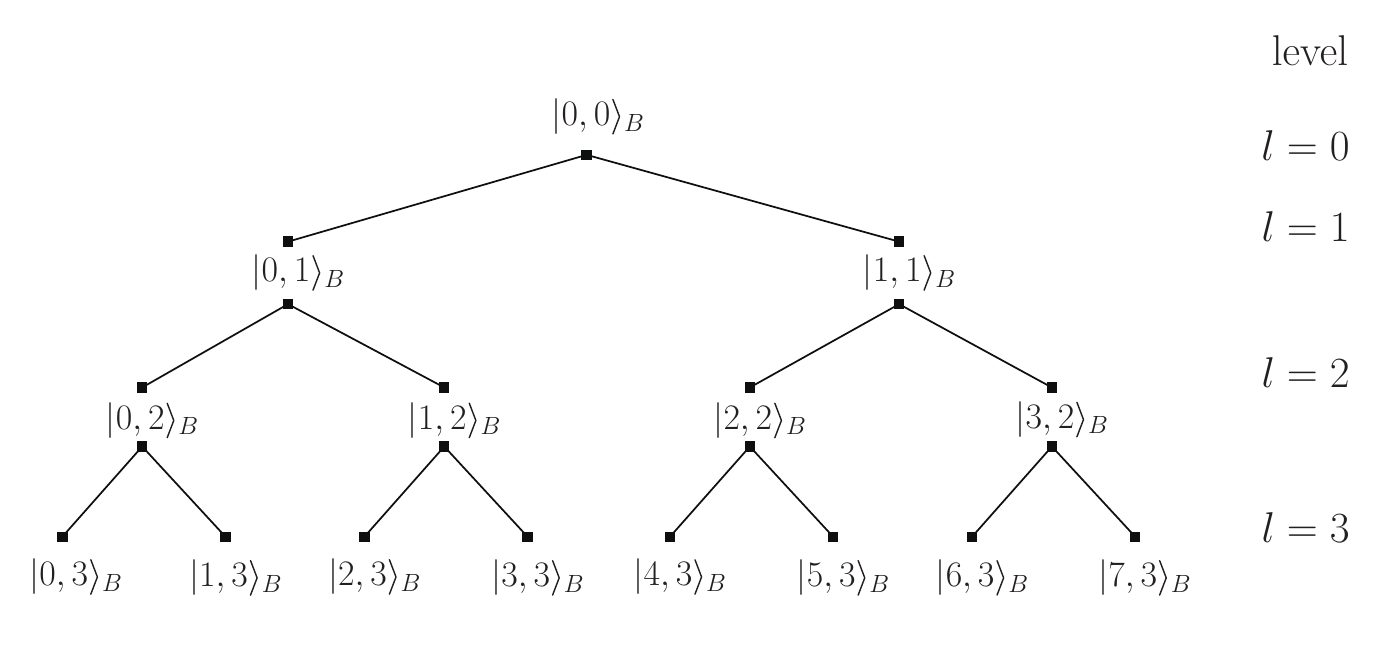}
\caption{A full binary tree with depth $n=3$. The state $|w,l\ket_B$ ($0\le w \le 2^l-1$,
$0\le l \le n$) denotes the position of the $w$th node counting from the left in level $l$.}
\label{tree}
\end{figure}
We consider the full binary tree with depth $n$ (i.e. it has
totally $2^n$ leaves corresponding to the memory cells) (see Fig.~\ref{tree} for $n=3$).
Let $|w,l\ket_B$ denote the position of 
the $w$th node counting from the left in level $l$ ($0\le w \le 2^l-1$,
$0\le l\le n$). We call  $V_B=\bigoplus_{w,l} V_{(w,l)}$ ($V_{(w,l)}=\mathbb{C}$)
spanned by $|w,l\ket_B$  ($0\le w \le 2^l-1$,
$0\le l \le n$) and
$V_C=\mathbb{C}^2$ spanned by  $|c\ket_C$ ($c\in\{0,1\}$) 
``bus space" and ``chirality space", respectively. The quantum walk
is defined on the space $V_B\otimes V_C$. The quantum walker at
the node $|w,l\ket_B$ moves to the left (resp. right) 
child node $|2w,l+1\ket_B$ (resp. $|2w+1,l+1\ket_B$) when the chirality of the walker is 
$|0\ket_C$ (resp. $|1\ket_C$).
This can be represented by the operator $\mathcal{S}_{(w,l)}$ acting on the space 
$(V_{(w,l)}\oplus V_{(2w,l+1)}\oplus V_{(2w+1,l+1)})\otimes V_C=\mathbb{C}^6$:
\begin{align}
&\mathcal{S}_{(w,l)}:\,\, |w,l\ket_B|0\ket_C\mapsto  |2w,l+1\ket_B|0\ket_C, \nn \\
&\mathcal{S}_{(w,l)}:\,\, |w,l\ket_B|1\ket_C\mapsto  |2w+1,l+1\ket_B|1\ket_C. 
\label{shift-d}
\end{align}
In addition, the walker at the left (resp. right) child node $|2w,l+1\ket_B$ 
(resp. $|2w+1,l+1\ket_B$) can be
pulled back to the parent node  $|w,l\ket_B$ by $\mathcal{S}_{(w,l)}$ only if its chirality is
$|0\ket_C$ (resp. $|1\ket_C$), and stays at the present position if its chirality is
$|1\ket_C$ (resp. $|0\ket_C$). Explicitly,
\begin{align}
&\mathcal{S}_{(w,l)}:\,\, |2w,l+1\ket_B|0\ket_C\mapsto  |w,l\ket_B|0\ket_C,\nn \\
&\mathcal{S}_{(w,l)}:\,\, |2w+1,l+1\ket_B|1\ket_C\mapsto  |w,l\ket_B|1\ket_C, \nn \\
&\mathcal{S}_{(w,l)}:\,\, |2w,l+1\ket_B|1\ket_C\mapsto  |2w,l+1\ket_B|1\ket_C,\nn \\
&\mathcal{S}_{(w,l)}:\,\, |2w+1,l+1\ket_B|0\ket_C\mapsto  |2w+1,l+1\ket_B|0\ket_C.
\label{shift-u}
\end{align}
Namely, $\mathcal{S}_{(w,l)}\in 
\End((V_{(w,l)}\oplus V_{(2w,l+1)}\oplus V_{(2w+1,l+1)})\otimes V_C)$ 
is a unitary operator expressed as
\begin{align}
\mathcal{S}_{(w,l)}=&
\sum_{i=0}^1\Bigl(
\left|2w+i,l+1\right\ket \left\bra w,l\right|
+
\left|w,l\right\ket\left\bra 2w+i,l+1\right|
\Bigr)_B\otimes  |i\ket\bra i|_C\nn \\
&+\sum_{i=0}^1 
\left|2 w+\frac{1+(-1)^i}{2},l+1\right\ket
\left\bra
2 w+\frac{1+(-1)^i}{2},l+1\right|_B
\otimes 
\left|i\right\ket\left\bra i\right|
_C.
\label{shift}
\end{align}

Combining a unitary operator $\mathcal{C}$ acting on $V_C$, we obtain
a non-trivial quantum motion on the graph. In the next section,
we construct $\mathcal{C}$  such that it acts on both 
$V_C$ and the ``address space" $V_A=({\mathbb{C}^2})^{\otimes n}$
to move the quantum walker (bucket) from the root to a specific leaf (memory cell)
and to return the bucket filled with information back to the original root.

%

%%%%%%%%%%%%%%%%%%%%%%%%%%%%%%%%%%%%%%%%%%%%%%%%%%%%%%%%%%%%
\section{qRAM via quantum walk}
%%%%%%%%%%%%%%%%%%%%%%%%%%%%%%%%%%%%%%%%%%%%%%%%%%%%%%%%%%%%
%
%
\begin{figure}[ttt]
\centering
\includegraphics[width=0.8\textwidth]{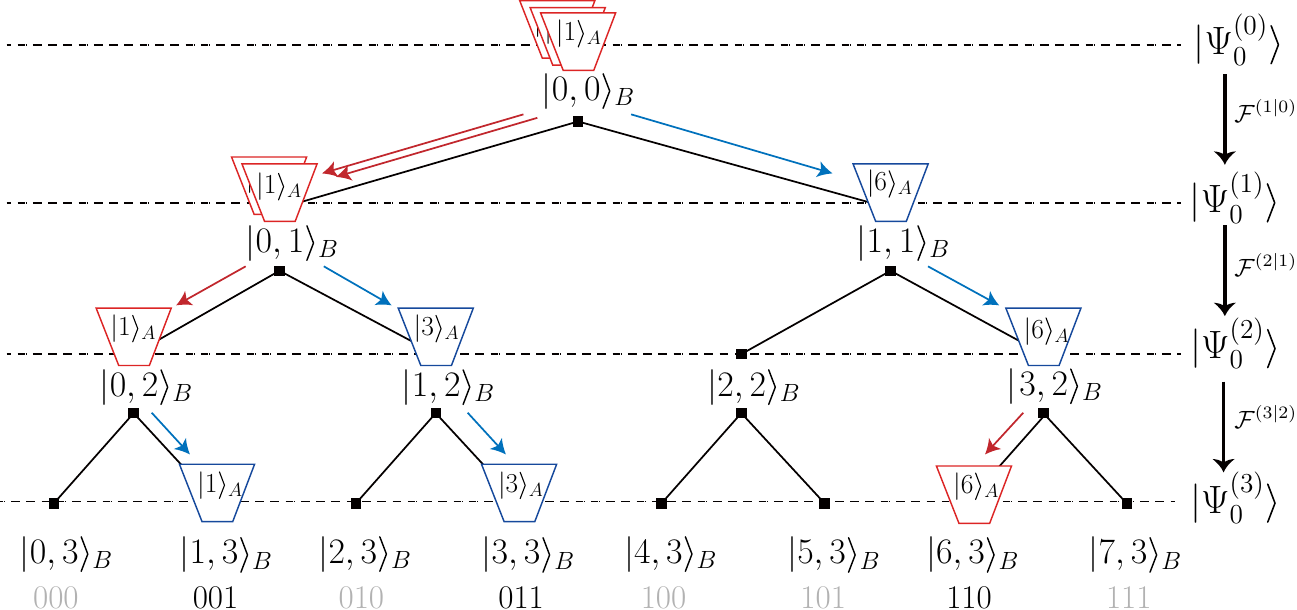}
\caption{A pictorial example of the routing scheme \eqref{routing-scheme}
for the case that $n=3$ and the input addresses are
$\sum|a\ket_A=
|001\ket_A+|011\ket_A+|110\ket_A$. 
Namely, the initial state  is given 
by $|\Phi_0^{(0)}\ket=\sum_a |0,0\ket_B |0\ket_C|a\ket_A|0\ket_D$, and depicted 
as the red bucket (in quantum superposition) located at the node $|0,0\ket_B$. Red 
corresponds to the chirality $|0\ket_C$ and blue to $|1\ket_C$.
Operating $\mathcal{F}^{(1|0)}$
defined by \eqref{F-matrix} on  $|\Phi_0^{(0)}\ket$, one has
$|\Phi_0^{(1)}\ket=|0,1\ket_B|0\ket_C(|001\ket_A+|011\ket_A)|0\ket_D+
|1,1\ket_B|1\ket_C|110\ket_A|0\ket_D$. Similarly, $\mathcal{F}^{(2|1)}|\Phi_0^{(1)}\ket$
yields $|\Phi_0^{(2)}\ket=|0,2\ket_B|0\ket_C|001\ket_A|0\ket_D+
|1,2\ket_B|1\ket_C|011\ket_A|0\ket_D+|3,2\ket_B|1\ket_C|110\ket_A|0\ket_D.$
Finally, $\mathcal{F}^{(3|2)}|\Phi^{(2)}_0\ket$ reads
$\Phi^{(3)}=|1,3\ket_B|1\ket_C|001\ket_A|0\ket_D+
|3,3\ket_B|1\ket_C|011\ket_A|0\ket_D+|6,3\ket_B|0\ket_C|110\ket_A|0\ket_D$.
Thus, one can deliver the bucket in superposition to the designated addresses $\sum |a\ket_A$.
}
\label{routing-pic}
\end{figure}
To appropriately access and retrieve data stored in the specified 
memory cells in the form of quantum superposition, we employ
the quantum walk explained in the previous section. 
Let 
\begin{align}
&|x^{(a)} \ket_D=|x^{(a)}_{m-1}\cdots x^{(a)}_{0}\ket_D=
|x^{(a)}_{m-1}\ket_{D_{m-1}}
\cdots |x^{(a)}_{0}\ket_{D_0}   \quad (x^{(a)}_i \in \{0,1\};\, 0\le i\le m-1 ), \nn \\
&|x^{(a)}_i\ket\in V_{D_i}=\mathbb{C}^2,\quad 
|x^{(a)}\ket_D\in V_D=\bigotimes_{i=0}^{m-1}V_{D_i}=(\mathbb{C}^2)^{\otimes m}
\end{align}
be the binary representation of data stored in a memory cell. Here, $|a\ket_A$
($0\le a \le N-1$; $N=2^n$) represents the address of the cell: 
\begin{align}
&|a\ket_A=|a_{n-1}\cdots a_0\ket_{A} 
=|a_{n-1}\ket_{A_{n-1}}\cdots |a_{0}\ket_{A_0} 
\quad (a_i\in \{0,1\}; \, 0\le i\le n-1),\nn \\
&|a_i\ket\in V_{A_i}=\mathbb{C}^2, \quad |a\ket_A\in V_A
=\bigotimes_{i=0}^{n-1}V_{A_i}=(\mathbb{C}^2)^{\otimes n}.
\end{align}
Let us call $V_A$ and $V_D$ the ``address space" and the ``data space",
respectively.
Our qRAM is defined on the space 
\begin{equation}
V:=
V_B\otimes V_C\otimes V_A\otimes V_D
\end{equation}
as
\begin{equation}
\text{qRAM}:\,\, \sum_{a\in \mathscr{A}} |0,0\ket_B |0\ket_C |a\ket_A
|0\ket_D \mapsto 
\sum_{a\in \mathscr{A}}
|0,0\ket_B|0\ket_C|a\ket_A|x^{(a)} \ket_D, 
\label{qRAM-def}
\end{equation}
where $\mathscr{A}\subset \{0,1,\cdots,N-1\}$.
\begin{figure}[ttt]
\centering
\includegraphics[width=0.8\textwidth]{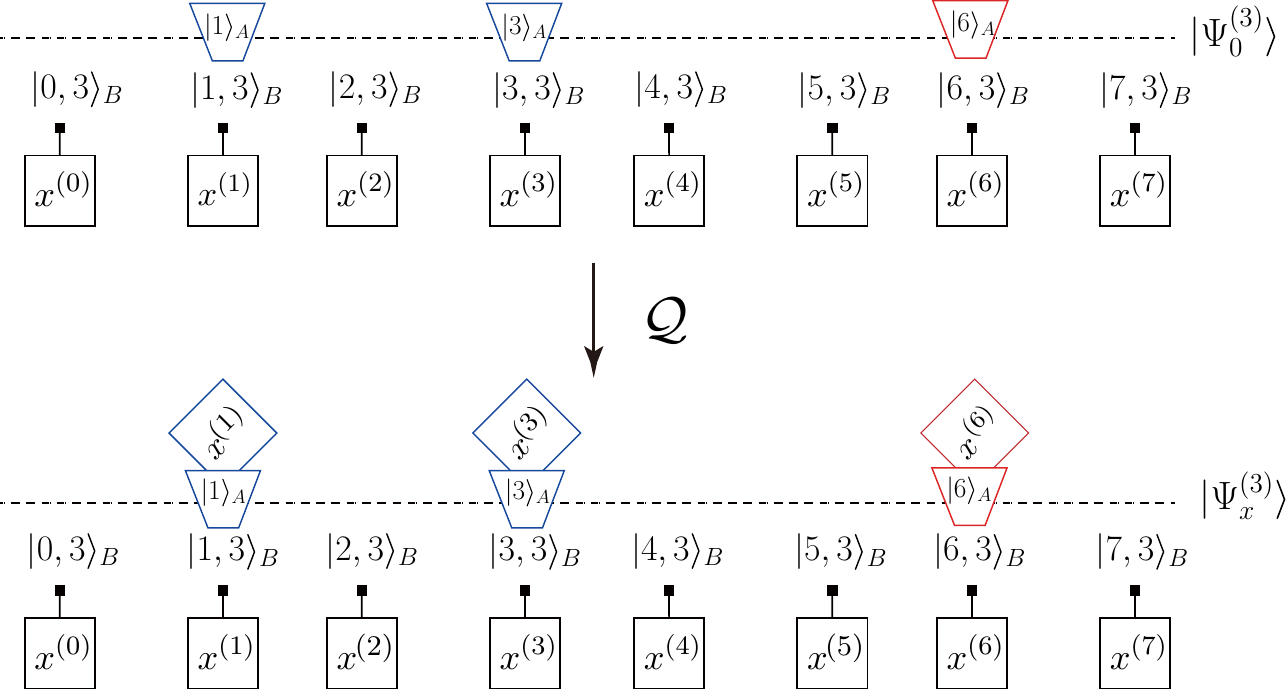}
\caption{A pictorial example of the querying scheme
\eqref{querying-scheme} and \eqref{querying-scheme2}
for the empty bucket (in quantum superposition) delivered at the memory cells 
addressed by $\sum|a\ket_A=|001\ket_A+|011\ket_A+|110\ket_A$ (see Fig.~\ref{routing-pic}
for  $|\Phi^{(3)}_0\ket$). Performing $\mathcal{Q}$ on the state $|\Phi^{(3)}_0\ket$,
one obtains $|\Phi^{(3)}_x\ket=|1,3\ket_B|1\ket_C|001\ket_A|x^{(1)}\ket_D+
|3,3\ket_B|1\ket_C|011\ket_A|x^{(3)}\ket_D+|6,3\ket_B|0\ket_C|110\ket_A|x^{(6)}\ket_D$
which denotes the bucket filled with the data.} 
\label{querying-pic}
\end{figure}

As described in Sec.~1, qRAM consists of the following three schemes: (i) a
routing scheme to move the empty bucket in a superposition
to specific cells, (ii)
a querying scheme to fill the bucket with data and (iii)
an output scheme to pull back the bucket and output the
data in the form of quantum superposition states. Correspondingly,
the qRAM is decomposed into the following three operators:
\begin{equation}
\text{qRAM}=\mathcal{F}^{\dagger} \mathcal{Q} \mathcal{F}\in \End(V).
\label{qRAM-op}
\end{equation}

\noindent
(i) {\it Routing scheme}. First we construct the routing scheme $\mathcal{F}\in \End(V_A\otimes
V_B\otimes V_C)$.
Let $|\Psi_0^{(l)}\ket \in V$ be a state that the bucket in a superposition 
is located at nodes in level $l$ (see Fig.~\ref{routing-pic}  as an example).
We set 
\begin{equation}
|\Psi_0^{(0)}\ket=\sum_{a\in\mathscr{A}}|0,0\ket_B |0\ket_C |a\ket_A |0\ket_D
\end{equation}
as the initial state. One finds that the bucket in a superposition is appropriately delivered
to the desired cells at $\sum_{a\in\mathscr{A}} |a\ket_A$ by $\mathcal{F}$:
\begin{equation}
\mathcal{F}:\,\, |\Psi_0^{(0)}\ket\mapsto|\Psi_0^{(n)}\ket,
\label{routing-scheme}
\end{equation}
which is  decomposed into
$
\mathcal{F}=\mathcal{F}^{(n|n-1)}\cdots \mathcal{F}^{(1|0)},
$
where the element $\mathcal{F}^{(l+1|l)}$
\begin{equation}
\mathcal{F}^{(l+1|l)}:\,\, |\Psi_0^{(l)}\ket\mapsto|\Psi_0^{(l+1)}\ket,
\quad
|\Psi_0^{(l)}\ket:=\left|\sum_{k=1}^l2^{l-k} a_{n-k},l \right\ket_B
|a_{n-l}\ket_C|a\ket_A |0\ket_D
\label{F-matrix}
\end{equation}
is given by
\begin{equation}
\mathcal{F}^{(l+1|l)}:=\sum_{w=0}^{2^l-1}
\mathcal{S}_{(w,l)}\mathcal{C}_{C,A_{n-(l+1)}}\mathcal{C}_{C,A_{n-l}}.
\end{equation}
Here, $\mathcal{S}_{(w,l)}$ is the shift operator defined by \eqref{shift} (see also 
\eqref{shift-d} and \eqref{shift-u}) and $\mathcal{C}_{C,{A_l}}\in \End(V_C\otimes V_{A_l})$ 
is  so-called the controlled NOT operator
defined as
\begin{equation}
\mathcal{C}_{C,A_l}:=I_C\otimes |0\ket\bra 0|_{A_l}+
X_C\otimes|1\ket\bra 1|_{A_l}, \quad \mathcal{C}_{C,A_n}:=I_C,
\label{C-op}
\end{equation}
where $X_C, I_C\in \End(V_C)$ are, respectively, the Pauli $X$ operator and
the identity matrix. Eq.~\eqref{F-matrix} can be recursively derived as follows: 
\begin{align}
&|\Psi_0^{(0)}\ket=\sum_{a\in\mathscr{A}}|0,0\ket_B |0\ket_C |a\ket_A |0\ket_D 
\xrightarrow{\mathcal{F}^{(1|0)}=\mathcal{S}_{(0,0)}\mathcal{C}_{C,A_{n-1}}\mathcal{C}_{C,A_{n}}}
\nn \\
&|\Psi_0^{(1)}\ket=
\sum_{a\in\mathscr{A}}|a_{n-1},1\ket_B |a_{n-1}\ket_C |a\ket_A |0\ket_D 
\xrightarrow{\mathcal{F}^{(2|1)}=(\mathcal{S}_{(0,1)}+\mathcal{S}_{(1,1)})
\mathcal{C}_{C,A_{n-2}}\mathcal{C}_{C,A_{n-1}}}\nn \\
&
|\Psi_0^{(2)}\ket=
\sum_{a\in\mathscr{A}}|2a_{n-1}+ a_{n-2},2\ket_B |a_{n-2}\ket_C |a\ket_A |0\ket_D 
\cdots
\xrightarrow{\mathcal{F}^{(l|l-1)}=\sum_{w=0}^{2^{l-1}-1}\mathcal{S}_{(w,l-1)}
\mathcal{C}_{C,A_{n-l}}\mathcal{C}_{C,A_{n-(l-1)}}}\nn \\
&
|\Psi_0^{(l)}\ket=
\sum_{a\in\mathscr{A}} \left|\sum_{k=1}^l 2^{l-k}a_{n-k},l\right\ket_B |a_{n-l}\ket_C |a\ket_A |0\ket_D 
\cdots 
\xrightarrow{\mathcal{F}^{(n|n-1)}=\sum_{w=0}^{2^{n-1}-1}\mathcal{S}_{(w,n-1)}
\mathcal{C}_{C,A_{0}}\mathcal{C}_{C,A_{1}}}\nn \\
&
|\Psi_0^{(n)}\ket=
\sum_{a\in\mathscr{A}} |a,n\ket_B |a_{0}\ket_C |a\ket_A |0\ket_D. 
\label{AAA}
\end{align}
Note that, in the last step, we have used
$a=\sum_{k=1}^n 2^{n-k} a_{n-k}$. As the result, the bucket in a superposition state
is delivered to the desired memory cells located at $\sum_{a\in\mathscr{A}}|a\ket$.
In Fig.~\ref{routing-pic}, we pictorially show the routing scheme. 
One easily sees that $O(n)$ steps are required for the routing scheme.
\\

\begin{figure}[t]
\centering
\includegraphics[width=0.8\textwidth]{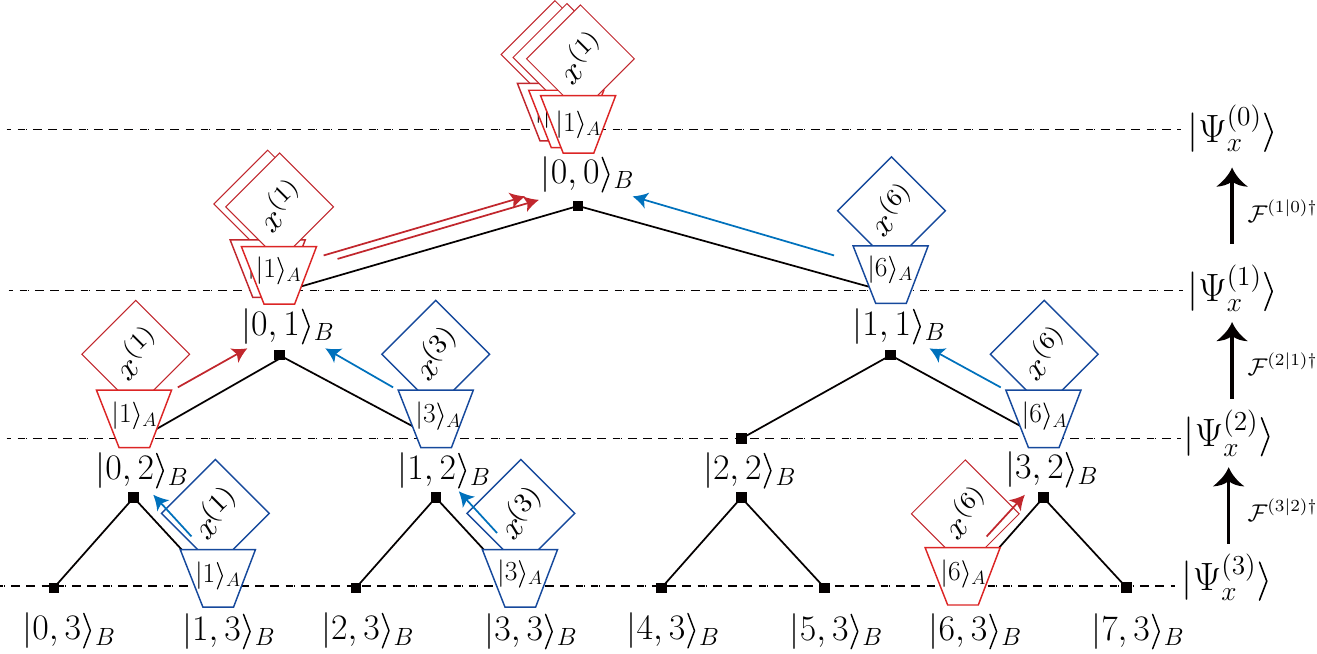}
\caption{A pictorial example of the output scheme
\eqref{output-scheme}. Due to the unitarity of $\mathcal{F}$ \eqref{routing-scheme},  
the bucket filled with the data (see Figs.~\ref{routing-pic} and
\ref{querying-pic}) can be pulled back to the root in the opposite procedure as the routing scheme:
$|\Phi^{(0)}_x\ket=\mathcal{F}^{\dagger}|\Phi^{(3)}_x\ket=
\sum_a|0,0\ket_B|0\ket_C|a\ket_A|x^{(a)}\ket$.}
\label{output-pic}
\end{figure}
\noindent
(ii) {\it Querying scheme}.
Since our schemes are independent of each other, 
the querying scheme $\mathcal{Q}\in \End(V_B\otimes V_D)$ to retrieve information from the 
memory cells 
can be significantly simplified and easily parallelized. The operator $\mathcal{Q}$  defined as
\begin{equation}
\mathcal{Q}:\,\, |\Psi_0^{(n)}\ket \mapsto  |\Psi_x^{(n)}\ket,
\quad
|\Psi_x^{(n)}\ket:=
\sum_{a\in\mathscr{A}} |a,n\ket_B|a_{0}\ket_C |a\ket_A |x^{(a)}\ket_D
\label{querying-scheme}
\end{equation}
can be simply composed by the Pauli $X$ operator $X_{D_i}\in\End(V_{D_i})$:
\begin{equation}
\mathcal{Q}:=\sum_{a=0}^{2^n-1}|a,n\ket\bra a,n|_B\otimes
\left[\bigotimes_{i=0}^{m-1} \left(X_{D_i}\right)^{x_i^{(a)}}\right].
\label{querying-scheme2}
\end{equation}
The querying scheme is shown in Fig.~\ref{querying-pic}. Note that the time steps  necessary 
for the querying scheme is only $O(1)$.\\

\noindent
(iii)
{\it Output scheme}.
Due to the unitarity of the operators (see \eqref{shift-u} and \eqref{C-op}), 
one easily finds that the quantum walk is reversible.
Namely, the output scheme to pull back the bucket filled with the data can be achieved
in exactly the opposite manner as the routing scheme:
\begin{equation}
\mathcal{F}^{\dagger}:\,\, |\Psi_x^{(n)}\ket\mapsto|\Psi_x^{(0)}\ket,
\quad
|\Psi_x^{(0)}\ket:=
\sum_{a\in\mathscr{A}} |0,0\ket_B |0\ket_C |a\ket_A |x^{(a)}\ket_D.
\label{output-scheme}
\end{equation}
See Fig.~\ref{output-pic} as an example of the output scheme.

Thus, we find that our qRAM \eqref{qRAM-op} satisfies \eqref{qRAM-def}.
The total steps required in our qRAM is $O(n)$ per memory call.

Here, let us briefly discuss the robustness against decoherence.
As introduced in Sec.~1, the GLM architecture must place
quantum devices (qutrits) at all the nodes (see Fig.~\ref{bucket}): totally
$O(n2^n)$ qutrits should be installed for the routing scheme. To route to
the memory cell, one activates the $O(n)$ qutrits at the nodes on the route. 
To lead the bucket to the desired memory cell and pull back the bucket filled with
the data to the root correctly, one must maintain the coherence
all the activated $O(n)$ qutrits (entangled with the $n$ address bits and $m$ data bits)
throughout all the schemes. In contrast, our qRAM does not
need any quantum devices at the nodes but relies on only one chirality 
state to essentially determine the course: the chirality state of the bucket is updated 
the moment the bucket passes each node on the route, as shown in \eqref{AAA}.
Namely, it only needs to maintain the coherence of the one chirality state
(entangled with the $n$ address bits and $m$ data bits)  for the short period
during the transfer of the bucket between the adjacent nodes. 
As a result, the cost to maintain the coherence can be 
significantly reduced.

\section{Toward a physical implementation}
In the previous section, we develop an algorithm of qRAM by 
quantum walk.
 The essential point in our qRAM architecture is the use of a quantum walk:
a quantum motion of a particle  with chirality. 

One might suspect that our qRAM algorithm \eqref{qRAM-def} 
presented in the previous section might be implemented as a quantum 
circuit without introducing quantum walks. Of course, it is
possible, since \eqref{qRAM-def} is defined as a combination of 
unitary operators. In that case, however, an extra quantum
manipulation is required to transfer the value of the address bits, 
one by one, to a qubit representing the chirality. In this sense, 
quantum walks are essential to the efficient implementation 
of our algorithm. The actual implementation of the qRAM intrinsically using a quantum walk
may be achieved by a method proposed in \cite{CGW}, where universal quantum 
computations via multi-particle quantum walks have been discussed. 
In the method \cite{CGW},
the so-called dual-rail encoding is employed:
the states $|0\ket$ and $|1\ket$ are expressed as a quantum particle  traveling
on two possible paths (see \eqref{rep-pic1} for example).
Single-qubit quantum gates can be implemented by some suitably constructed graphs
and single-particle scattering processes on them. On the other hand, two-qubit gates
are implemented by two-particle scattering on graphs. Namely, arbitrary unitary 
transformations can be represented as multi-particle scattering processes, which
could be used to design a quantum architecture without need for time-dependent 
control.

Here, we present  some specific examples of how to express the quantum states in 
our qRAM architecture using multiple quantum walkers. The actual implementation
of the operators such as $\mathcal{S}$ \eqref{shift}  and $\mathcal{C}$ 
\eqref{C-op} by some scattering processes of multiple quantum walkers 
is deferred to a future work.

\begin{figure}[t]
\centering
\includegraphics[width=0.95\textwidth]{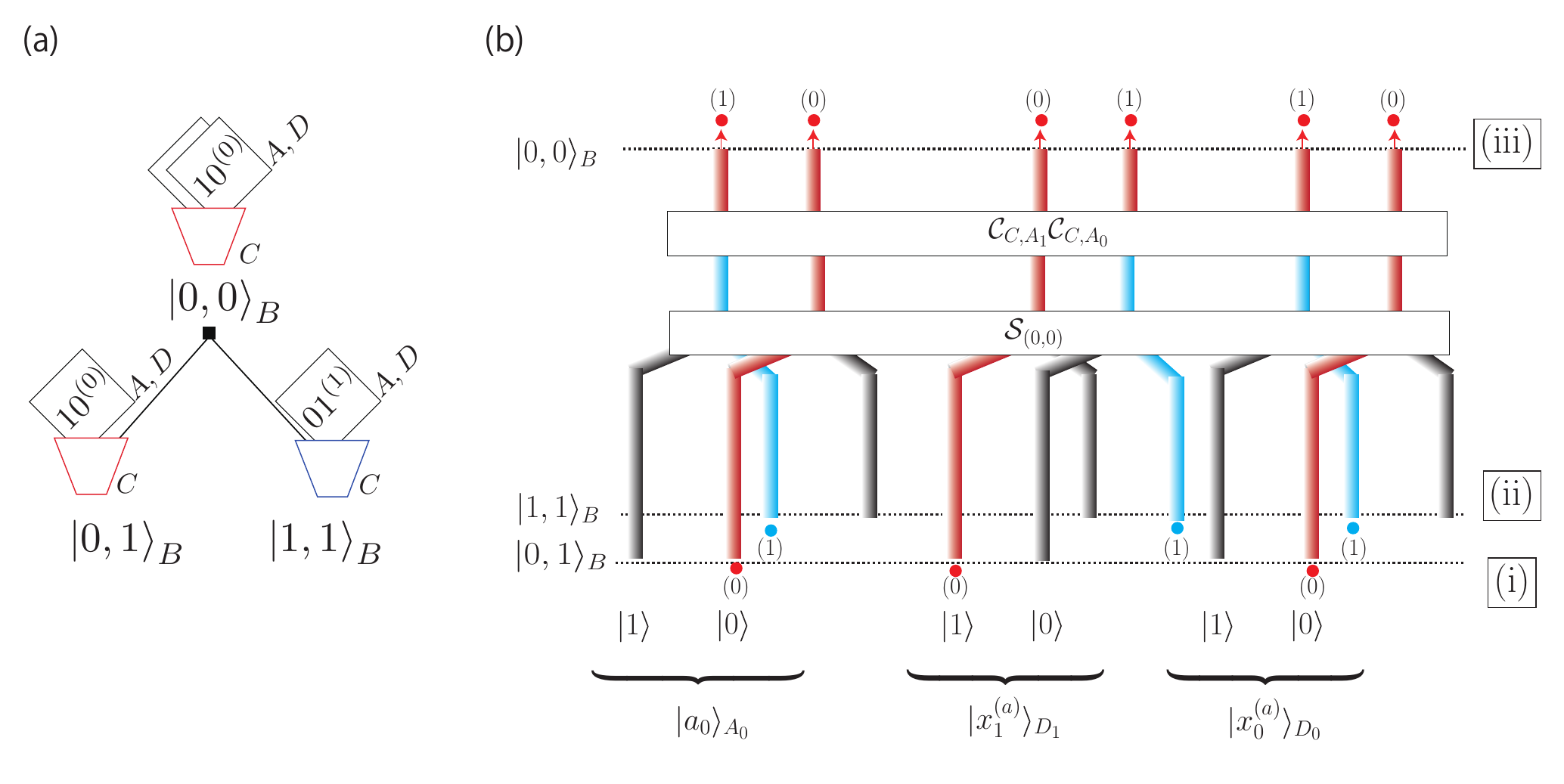}
\caption{(a): An example  of the output scheme $\mathcal{F}^{\dagger}$
\eqref{output-eg} at a node $|0,0\ket_B$ for $n=1$ and $m=2$.
(b): A representation of (a) by quantum walkers (see also 
\eqref{rep-pic1}--\eqref{rep-pic3}
as examples of the representation of states).
(i), (ii) and (iii) represents the 
states 
(i) $|0,1\ket_B|0\ket_C|0\ket_A|10^{(0)}\ket_D$,
(ii) $|1,1\ket_B|1\ket_C|1\ket_A|01^{(1)}\ket_D$
and (iii) 
$|0,0\ket_B|0\ket_C\left(|0\ket_A|10^{(0)}\ket_D+
|1\ket_A|01^{(1)}\ket_D\right)$, respectively. Note that 
the actual implementation of the operators $\mathcal{S}$ \eqref{shift}  and 
$\mathcal{C}$ \eqref{C-op} by some scattering processes
of multiple quantum walkers on properly constructed rails
is left as a future task.}
\label{rep-pic4}
\end{figure}

In total,   $n+m$ quantum walkers with chirality $c$ passing through $2(n+m)$ different 
``rails" represent a state $|c\ket_C|a\ket_A\ket|x^{(a)}\ket_D$. For instance,
an arbitrary state  $|y\ket_{A,D}=|y_{n+m-1}
\cdots y_0\ket_{A,D}$ ($y_i\in \{0,1\}$; $0\le i\le n+m-1$)
with chirality $|c\ket_C$ can be represented as
\begin{equation}
\includegraphics{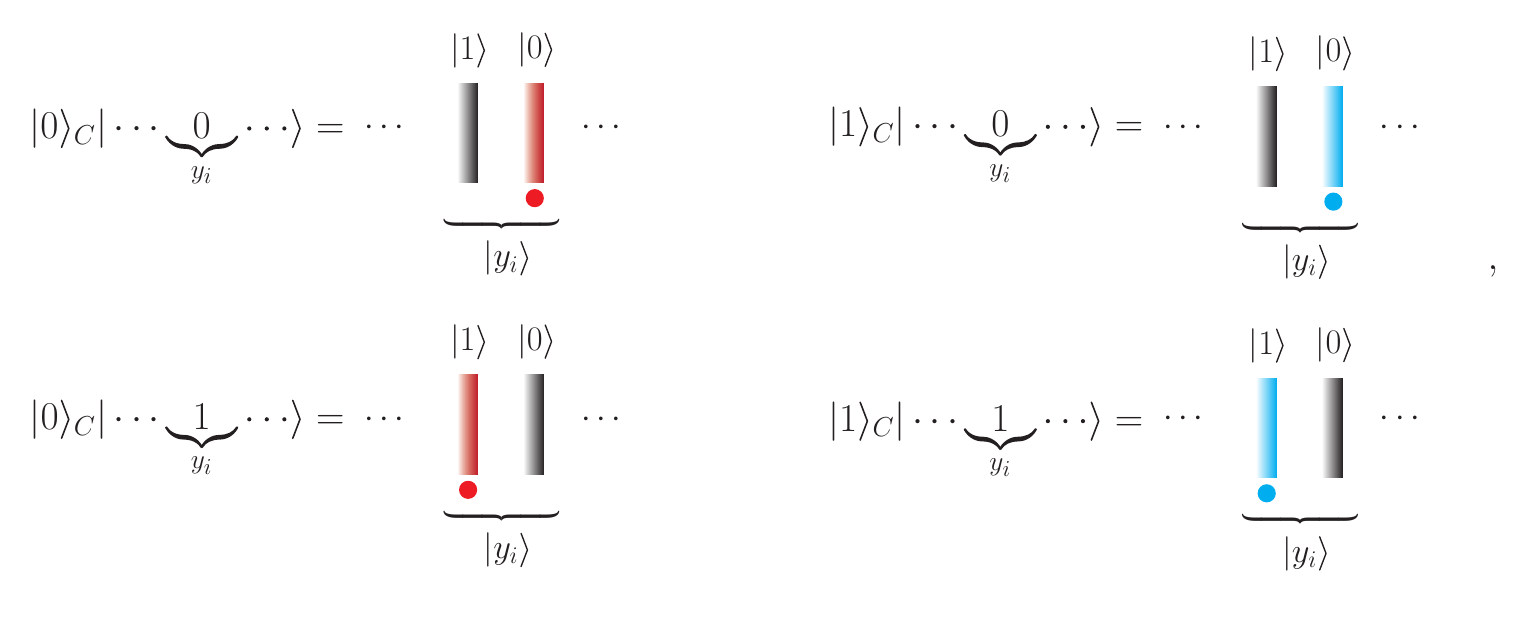}
\label{rep-pic1}
\end{equation}
where the walker with chirality $|0\ket_C$ (resp. $|1\ket_C$) is expressed as the rail 
colored red (resp. blue). 
The superposition states are also characterized by, for instance,
\begin{equation}
\includegraphics{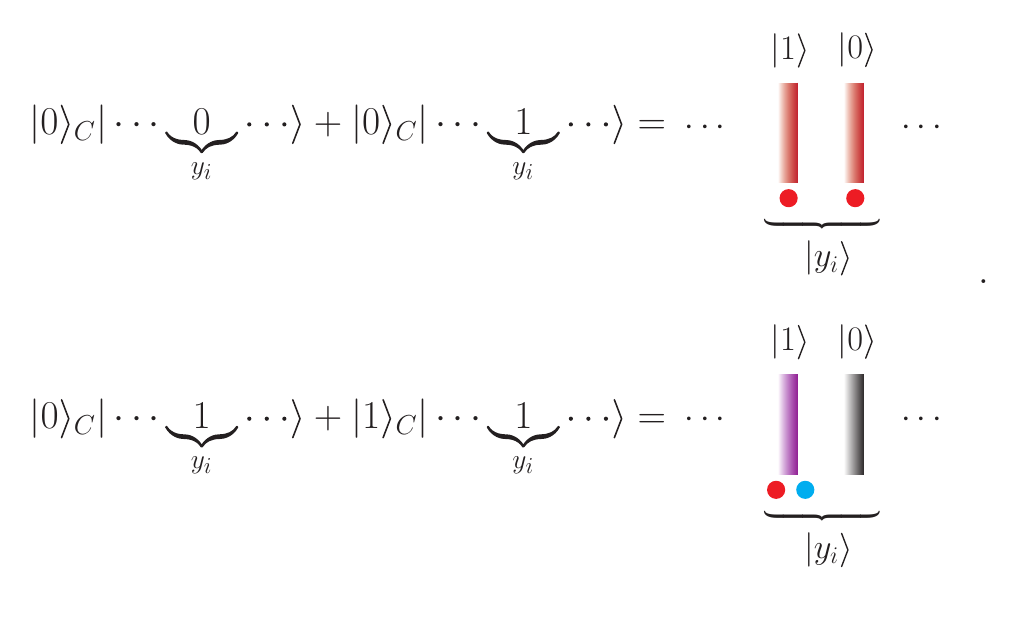}
\label{rep-pic2}
\end{equation}
As a more complicated example, we give
\begin{equation}
\includegraphics{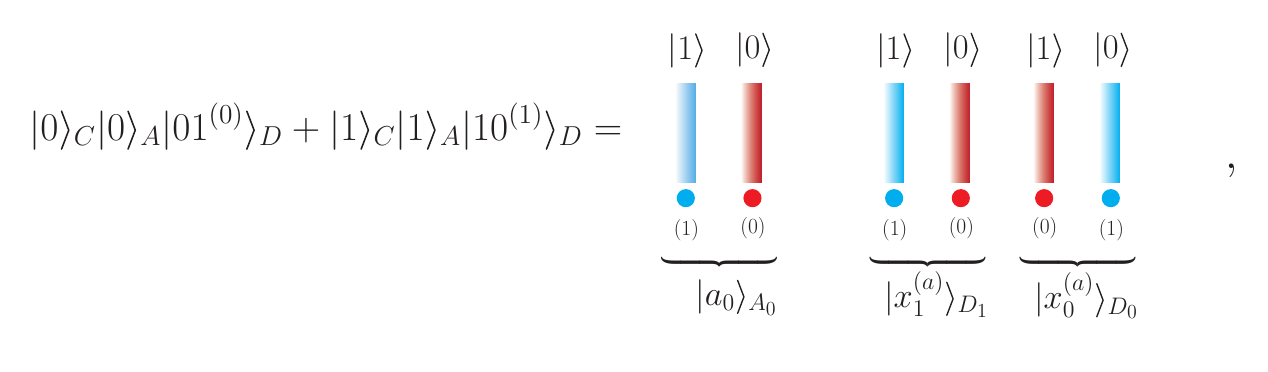}
\label{rep-pic3}
\end{equation}
where $(0)$ and (resp. $(1)$) in the r.h.s. denotes the correlation
with the particle representing the address $a_0=0$ (resp. $a_0=1$).
Thus the state $\Psi_0^{(l)}$ or $\Psi_x^{(l)}$ can be expressed by
$n+m$ quantum walkers and their superpositions passing through the 
$2(n+m)\times 2^{l}$ rails. In Fig~\ref{rep-pic4},  
we depict an example 
of the output scheme \eqref{output-scheme} 
for  $n=1$ and $m=2$.
\begin{align}
\mathcal{F}^{\dagger}:\,\,
&|0,1\ket_B|0\ket_C|0\ket_A|10^{(0)}\ket_D+
|1,1\ket_B|1\ket_C|1\ket_A|01^{(1)}\ket_D\nn \\
&\qquad \mapsto
|0,0\ket_B|0\ket_C\left(|0\ket_A|10^{(0)}\ket_D+
|1\ket_A|01^{(1)}\ket_D\right).
\label{output-eg}
\end{align}

Our proposal for qRAM  might be useful to design a qRAM architecture
with simpler structures due to the following reasons. First, the routing and output 
scheme might be represented as some multi-particle scattering processes on  suitably 
constructed graphs at each node, and therefore our bucket brigade qRAM does not need any quantum
devices (e.g. qudits) at 
the nodes, but only needs $m+n$ quantum particles as quantum resources.
Second, the quantum walkers are not entangled with the nodes. Thus, the three schemes, 
i.e. the routing,
querying and output schemes can be achieved completely independently, and hence a full 
parallelization can be easily accomplished.
This is in contrast to the original bucket brigade scheme \cite{GLM-PRL,GLM08}, where
we must maintain the coherence between the bucket and the $O(n)$ nodes on
the route: some ingenuity needs for a fully parallelization (see \cite{POB} for instance).
Finally, since  the three schemes can be described as  dynamics of the quantum walkers,
we could design qRAM with no need of time-dependent control.
%%%%%%%%%%%%%%%%%%%%%%%%%%%%%%%%%%%%%%%%%%%%%%%%%%%%%%%%%%%%
\section{Concluding remarks}
%%%%%%%%%%%%%%%%%%%%%%%%%%%%%%%%%%%%%%%%%%%%%%%%%%%%%%%%%%%%
%
In summary, we have provided a new concept of bucket brigade qRAM utilizing a 
quantum walk. Controlling a quantum motion of the quantum bucket with chirality,
we can efficiently deliver the bucket to the  desired memory cell,
and fill the bucket with data stored in cells. Since our qRAM does
not rely on  {\it quantum} switches for the routing scheme, the bucket
is free from the entanglement with the nodes.
Alternatively, it uses the chirality state, which entangles with the address bits for a short amount of time, to navigate the quantum walker.
Our qRAM may be more robust against quantum decoherence compared to the conventional bucket brigade method where it is required to maintain quantum  coherence in all the activated qutrits.

The actual implementation of our algorithms by 
scattering processes of quantum walkers on properly constructed graphs 
\cite{CGW,Childs, LCETK} remains a future problem. 
It may also be of interest to apply our qRAM to quantum information 
processing, for instance image processing and transformations utilizing quantum
version of fast Fourier transform \cite{ASY}, where the generation
of multiple quantum images is crucial.
%
%%%%%%%%%%%%%%%%%%%%%%%%%%%%%%%%%%%%%%%%%%%%%%%%%%%%%%%%%%%%
\section*{Acknowledgment}
%%%%%%%%%%%%%%%%%%%%%%%%%%%%%%%%%%%%%%%%%%%%%%%%%%%%%%%%%%%%
%
The present work was partially supported by Grant-in-Aid for Scientific
Research (C) No. 20K03793 from the Japan Society for the 
Promotion of Science.

\end{document}